\documentclass[11pt]{article}
\usepackage{epsfig}
\usepackage{graphicx,amssymb,amsmath}
\usepackage{latexsym}
\usepackage[dvips]{color}
\begin{document}
\newcommand{\M}{\mbox{m}}
\newcommand{\n}{\mbox{$n_f$}}
\newcommand{\EP}{\mbox{$e^+$}}
\newcommand{\EM}{\mbox{$e^-$}}
\newcommand{\EPEM}{\mbox{$e^+e^{-}$}}
\newcommand{\EMEM}{\mbox{$e^-e^-$}}
\newcommand{\GG}{\mbox{$\gamma\gamma$}}
\newcommand{\GE}{\mbox{$\gamma e$}}
\newcommand{\GP}{\mbox{$\gamma e^+$}}
\newcommand{\TEV}{\mbox{TeV}}
\newcommand{\GEV}{\mbox{GeV}}
\newcommand{\LGG}{\mbox{$L_{\gamma\gamma}$}}
\newcommand{\LGE}{\mbox{$L_{\gamma e}$}}
\newcommand{\LEE}{\mbox{$L_{ee}$}}
\newcommand{\LEPEM}{\mbox{$L_{e^+e^-}$}}
\newcommand{\WGG}{\mbox{$W_{\gamma\gamma}$}}
\newcommand{\WGE}{\mbox{$W_{\gamma e}$}}
\newcommand{\EV}{\mbox{eV}}
\newcommand{\CM}{\mbox{cm}}
\newcommand{\MM}{\mbox{mm}}
\newcommand{\NM}{\mbox{nm}}
\newcommand{\MKM}{\mbox{$\mu$m}}
\newcommand{\SEC}{\mbox{s}}
\newcommand{\CMS}{\mbox{cm$^{-2}$s$^{-1}$}}
\newcommand{\MRAD}{\mbox{mrad}}
\newcommand{\IND}{\hspace*{\parindent}}
\newcommand{\E}{\mbox{$\epsilon$}}
\newcommand{\EN}{\mbox{$\epsilon_n$}}
\newcommand{\EI}{\mbox{$\epsilon_i$}}
\newcommand{\ENI}{\mbox{$\epsilon_{ni}$}}
\newcommand{\ENX}{\mbox{$\epsilon_{nx}$}}
\newcommand{\ENY}{\mbox{$\epsilon_{ny}$}}
\newcommand{\EX}{\mbox{$\epsilon_x$}}
\newcommand{\EY}{\mbox{$\epsilon_y$}}
\newcommand{\BI}{\mbox{$\beta_i$}}
\newcommand{\BX}{\mbox{$\beta_x$}}
\newcommand{\BY}{\mbox{$\beta_y$}}
\newcommand{\SX}{\mbox{$\sigma_x$}}
\newcommand{\SY}{\mbox{$\sigma_y$}}
\newcommand{\SZ}{\mbox{$\sigma_z$}}
\newcommand{\SI}{\mbox{$\sigma_i$}}
\newcommand{\SIP}{\mbox{$\sigma_i^{\prime}$}}
\newcommand{\be}{\begin{equation}}
\newcommand{\ee}{\end{equation}}
\newcommand{\bc}{\begin{center}}
\newcommand{\ec}{\end{center}}
\newcommand{\bi}{\begin{itemize}}
\newcommand{\ei}{\end{itemize}}
\newcommand{\ben}{\begin{enumerate}}
\newcommand{\een}{\end{enumerate}}
\newcommand{\bm}{\boldmath}

\title{\bf Photon colliders: The first 25 years~\thanks{Presented at
    PHOTON2005 conference, Aug. 30 - Sept.4, 2005, Warsaw, Poland}}
\author{V.I.~Telnov~\thanks{telnov@inp.nsk.su} \\[2mm]
{\it Budker Institute of Nuclear Physics, 630090 Novosibirsk, Russia} }
\date{} 
\maketitle
\begin{abstract}
  In this invited talk at the ``historical'' session of PHOTON2005, I
  was asked to recount the history and the development, from its
  earliest days to the present, of the idea of photon colliders based
  on conversion of high energy electrons to high-energy photons at a
  future high-energy linear \EPEM\ collider.  Described in this talk
  are the general features and schemes of a photon collider, the
  evolution in understanding of what the parameters of a realistic
  photon collider are, possible solutions of various technical
  problems, the physics motivation, and the present status of
  photon-collider development. For a more detailed description of the
  photon collider at the ILC and a discussion of the associated
  technical issues, please refer to my talks at PLC2005, the
  conference that immediately followed PHOTON2005 (to be published in
  Acta Physica Polonica B as well).
\end{abstract}

\section{Prehistory and the idea of the photon collider}

Photon colliders do not exist yet, but already have a rich 25-year
history.  The early history of \GG\ physics, studied mainly in
collisions of virtual photons at \EPEM\ storage rings, has been
presented at PHOTON2005 by S.~Brodsky~\cite{Brodsky} and
I.~Ginzburg~\cite{Ginzburg}.  Hence, I begin my narration by
describing the circumstances that led to the birth of the idea of the
high-energy photon collider. This is the first time I share an account
of these events with the public; this conference, subtitled ``The
Photon: its First Hundred Years and the Future,'' provides an
appropriate venue for such historical reviews.  I will also mention
the story of the observation of $C=+$ resonances in \GG\ collisions at
SLAC in 1979, which is also an important event in the \GG\ history.

Two-photon physics had been talked about since 1930s, but as an active
research field is began in early 1970s, when production of \EPEM\ 
pairs was discovered in collisions of {\bf virtual} photons at the
\EPEM\ storage ring VEPP-2 in Novosibirsk and theorists realized that
this method can be used to study a variety of two-photon processes.

To study two-photon physics at a greater depth, we in Novosibirsk
decided to build MD-1, a dedicated detector with a transverse magnetic
field and a tagging system for scattered electrons. Before experiments
at the VEPP-4 collider started, in 1978--79 I had the privilege of
having been able to visit SLAC for four months and work with the Mark
II group, where I observed two-photon production of the
$\eta^{\prime}$ and $f_2$ mesons. It then became clear that tagging of
the scattered electrons is not neccesary for study of many two-photon
processes; the MARK II paper on two-photon $\eta^{\prime}$
production~\cite{Abrams} triggered a wave of results from all \EPEM\ 
experiments.

Many interesting two-photon reactions were 
studied in the years that followed, but the results could not compete
with the revolutionary discoveries made in \EPEM\ annihilation. 
The reason for this is that the luminosity and the energy in virtual \GG\
collisions are small. Indeed, the number of equivalent photons
surrounding each electron is $dN_{\gamma}\sim 0.035 d\omega/\omega$, 
and the corresponding \GG\ luminosity for
$\WGG/2E_0>0.2$ is only $\LGG \approx 4\times 10^{-3} L_{e^+e^-}$,
which is an order of magnitude smaller than for $\WGG/2E_0>0.5$.

The other important element that led to the conception of the idea of
high-energy photon colliders is the activity on \EPEM\ linear
colliders in Novosibirsk. In December 1980, the First USSR workshop on
the physics at VLEPP was held in Novosibirsk~\cite{VLEPP1980}.  Only
one talk on \GG\ physics was on the agenda, an overview by I.~Ginzburg
and V.~Serbo of the physics of two-photon production of hadrons at
VLEPP energies (in collisions of virtual photons). I was not planning
to give any talks, but several days before the workshop began to think
about the possibility of converting electrons to \emph{real} photons
in order to increase the \GG\ luminosity at VLEPP. At the discussion
session, which was part of the workshop's schedule I gave a short talk
on this subject using blackboard.

The idea was rather simple. At linear colliders, electron beams are
used only once, which makes it possible to convert electrons to
photons, and thus to obtain collisions of real photons. All that is
needed is some sort of a target at a small distance from the
interaction point (IP), where the conversion would take place. For
example, if one were to place a target of $0.3 X_0$ thickness, the
number of bremsstrahlung photons would be greater than the number of
virtual photons by one order of magnitude, and the corresponding \GG\ 
luminosity would increase by two orders of magnitude; however, this
approach suffers from photo-nuclear backgrounds. I continued my talk
by saying that there are other methods of conversion: for example,
crystals are better than amorphous targets because the effective $X_0$
is much shorter, leading to smaller backgrounds; undulators produce
photons whose energies are too low\dots At this exact moment G.~Kotkin
interjected from his chair, ``Lasers\,!''. In fact, this method was
already well-known in our community: at SLAC, Compton backscattering
had been used since mid-1960s for production of high-energy photons;
in Novosibirsk, such a facility had been constructed for our
experiments at VEPP-4 for the measurement of the electron polarization
in the method of resonant beam depolarization.

During the following discussion, several people expressed quite a
negative reaction to the idea of laser $e \to \gamma$ conversion due
to the very low conversion probability.  In the 4.5-page summary on
two-photon physics written for the workshop proceedings by Ginzburg,
Serbo and me, there was only one paragraph about the photon-collider
idea, with the conclusion that a more detailed study is needed.

Immediately after this workshop, a group of \GG\ enthusiasts, namely:
I.~Ginzburg (Institute of Mathematics), G.~Kotkin (Novosibirsk State
University), V.~Serbo (also NSU) and V.~Telnov (INP) decided to pursue
the method of the laser photon conversion further: if feasible, it
would be the best among all the alternatives. It was a very exciting
study, and contributions from all members of this team were vitally
important to make possible the first publication and further advances
on the concept of photon colliders.

The method of production of high-energy photons by Compton scattering
of laser light off high-energy electrons was proposed in 1963 by
Arutyunian, Goldman and Tumanian~\cite{ARUT} and independently by 
Milburn~\cite{Milb}, and
soon afterwards was utilized~\cite{Kulikov,Ballam}. However, the
conversion coefficient was very small, about $k=N_{\gamma}/N_e \sim
10^{-7}$ \cite{Ballam}. For the photon collider, we needed $k \sim 1$,
seven orders of magnitude more\,!

We determined the required laser flash energy, then checked the literature on
powerful lasers, consulted with laser experts, and found that lasers
with required flash energies, about 10 J, already existed, albeit with
much longer pulse durations and lower repetition rates than those required
by a photon collider (the repetition rate in the VLEPP project circa 1980 
was only 10--100 Hz, compared with 15 kHz in the present ILC design).  
Discussions with laser experts gave us some
hope that these problems will be solved in future. Extrapolating
the progress of laser technologies into the next two decades and adding our
optimism, we came to the conclusion that a photon collider
based on laser photon conversion 
is not such a crazy idea and deserves being published.

The preprint INP 81-50, dated February 25, 1981 (in English), was sent
to all major HEP laboratories and to many individual physicists, but
publication of the corresponding paper was a problem. The original
submission of our paper to \emph{Pisma ZHETPh} was rejected:
\emph{``... the editorial board does not consider worthwhile a rapid
  publication of your article because the realization of such an
  experiment is not possible in the near future ...  lasers of the
  required parameters do not exist ...  and their creation is not
  likely in near future}.''  We resubmitted the article to the same
journal with additional comments, but once again received a
confirmation of the previous refusal. We then sent the paper to
\emph{Physics Letter}, were was declined as well, \emph{''...the
  article is very interesting but does not need urgent publication.
  You can publish it, for example, in Nuclear Instruments and
  Methods.''}  What do we do?  Fortunately, in August 1981 we had a
chance to meet personally with I.~Sobelman, the editor of \emph{Pisma
  ZHETPh}, who was visiting Novosibirsk; following that meeting, the
paper was published on November 5, 1981 (received March
10)~\cite{GKST81}. Two additional, more detailed papers written in
1981--1982~\cite{GKST83,GKST84} were published in NIM; their combined
citation index now surpasses 1000.

In September 1981, C.~Akerlof of the Univ. of Michigan published a
preprint~\cite{Akerlof} that contained a similar idea. However, he
considered only \GE\ collisions and underestimated the required laser
flash energy by 1--2 orders of magnitude. That was after two of our
preprints~\cite{GKST81,GKST83}, and mentioning of the photon-collider
concept in August 1981 at the Symposium On Lepton And Photon
Interactions At High Energies in Bonn in VLEPP status
report~\cite{Balakin}.  In November 1981, Kondratenko, Pakhtusova and
Saldin from our institute suggested the use of single-pass
free-electron lasers in a future photon collider~\cite{KONDR82}.

In the following sections, we consider the main principles and
features of photon colliders, the technical issues, and how the laser and
linear-collider technologies and our understanding of them evolved 
with time.

Two remarks are in order. Firstly, we consider only {\bf high-energy}
photon colliders of luminosities that are of real interest to particle
physics.  As for low-energy photon-photon scattering, in 1928--30
S.~Vavilov in attempted detection of scattering of visible photons
from two lamps~\cite{Vavilov}; later, people experimented with laser
photons, but these experiments also failed due to the very small cross
section for photon-photon scattering at low energies.  There existed
ideas of using synchrotron radiation, beamstrahlung photons, and even
nuclear explosions (Csonka~\cite{Csonka}) to study photon-photon
interactions.  Beamstrahlung photons can indeed have high energies,
but the idea is not practicable as collisions of virtual photons at
storage rings provide a much higher luminosity.

Secondly, it is well known that during collisions at \EPEM\ linear
colliders electrons and positrons emit hard photons, about one such
photon per electron.  So, simultaneously with \EPEM\ collisions, for
free, one gets a photon-photon collider of a high luminosity and a
rather high energy (typically several percents of the beam energy, but
can be higher).  At very high energies, the average energy of such
beamstrahlung photons is about 25\% of the electron energy. In 1988,
R.~Blankenbecler and S.~Drell even considered the prospects for such a
photon-photon collider in the quantum beamstrahlung
regime~\cite{Blankenbecler}. The disadvantages of this method are the
following~\cite{TEL90}: one needs a multi-TeV linear collider (or very
small beam sizes), the luminosity is limited by beam-collision
instabilities, the photon spectrum is wide, and in the strong field
($\Upsilon >1$) of the opposing beam the high-energy photons will
convert to \EPEM\ pairs. At the photon colliders based on Compton
backscattering, beamstrahlung photons contribute to the low-energy
part of the \GG\ luminosity spectrum and are taken into account in all
simulations.
\vspace{-0.2cm} \section{Nomenclature of linear-collider projects}
Over the past two decades, several projects of linear colliders were
in existence, see Table~\ref{tab}.  Only one of them, the SLC, was
actually built and operated quite successfully at $Z$-boson energy.
The SLC was quite a special linear collider, constructed on the base
of the existing SLAC linear accelerator by adding two arcs to achieve
\EPEM\ collisions.  Its luminosity was about 3 orders of magnitude
lower than the the luminosity that can be obtained at an optimized
linear collider.  At present, two projects remain: the International
Linear Collider (ILC) with the energy of up to 1 TeV and the Compact
LInear Collider (CLIC) with the energy of up to 3 (perhaps, 5) TeV.
Neither of the projects has been approved; however, there is an
``intent'', and a hope, to have the ILC built by 2015.

\begin{table}[!htb]
\vspace{-0.1cm}
\caption{Linear collider projects, the past and the present} \vspace{-0.1cm}
\bc
  {\renewcommand{\arraystretch}{1.15} \setlength{\tabcolsep}{2.mm}
\begin{tabular}{lllll} \hline \small
 Name & Center & Type &  Energy [GeV] & \;\;\;\;\;Years  \\ \hline
VLEPP & BINP   & $S$, $X$-band   &500--1000 & $\sim$ 1978--1995  \\
SLC   & SLAC  &  $S$-band  &  90 &$\sim$ 1987--2000 (oper.) \\
NLC   & SLAC  & $X$-band   & 500--1000   & $\sim$  1986--2004 \\
JLC   & KEK  &  $X$-band   & 500--1000   & $\sim$  1986--2004 \\
TESLA & DESY  & $L$-band, s-cond.  &  500--800  &$\sim$  1990--2004 \\
SBLC  &  DESY  & $S$-band  & 500--100   & $\sim$ 1992--1997 \\
CLIC  & CERN  & $X$, two-beam  &500--3000   &$\sim$ 1986--\dots  \\
ILC   & ????  & $L$-band, s-cond.  &  500--1000  & $\sim$ 2004--\dots \\
\end{tabular} \\
}
\ec
\label{tab}
\end{table} \vspace{-1.cm}

\section{Basics of the photon collider}

Here, we briefly consider the main characteristics of backward Compton
scattering and the requirements on the lasers.

\vspace{-0.2cm}
\subsection{Kinematics and photon spectra \label{basics}}
In the conversion region, a laser photon of energy $\omega_0$
collides with a high-energy electron of energy $E_0$ at a small
collision angle $\alpha_0$ (almost head-on).  The energy of the
scattered photon $\omega$ depends on the photon scattering angle
$\vartheta$ in respect to the initial direction of the electron as
follows~\cite{GKST83}: \vspace*{-1.3mm}

\begin{equation}
\omega = \frac{\omega_m}{1+(\vartheta/\vartheta_0)^2},\;\;\;\;
\omega_m=\frac{x}{x+1}E_0, \;\;\;\;
\vartheta_0= \frac{mc^2}{E_0} \sqrt{x+1},
\label{kin}
\end{equation} \vspace*{-0.8mm}

\noindent where  \vspace*{-0.9mm}

\begin{equation}
x=\frac{4E \omega_0 }{m^2c^4}\cos^2{\frac{\alpha_0}{2}}
 \simeq 15.3\left[\frac{E_0}{\TEV}\right]
\left[\frac{\omega_0}{eV}\right] =
 19\left[\frac{E_0}{\TEV}\right]
\left[\frac{\mu \mbox{m}}{\lambda}\right], 
\label{e3:x}
\end{equation}

\noindent $\omega_m$ being the maximum energy of scattered photons.
For example: $E_0 = 250$ GeV, $\omega_0 = 1.17$ eV ($\lambda=1.06$
\MKM) (for the most powerful solid-state lasers) $\Rightarrow$ $x=4.5$
and $\omega_m/E_0 = 0.82$.  Formulae for the Compton cross section can
be found elsewhere~\cite{GKST83,GKST84}.

\begin{figure}[!htb]
\begin{minipage}{0.45\linewidth}
\hspace{-0.cm}\includegraphics[width=7.5cm,angle=0]{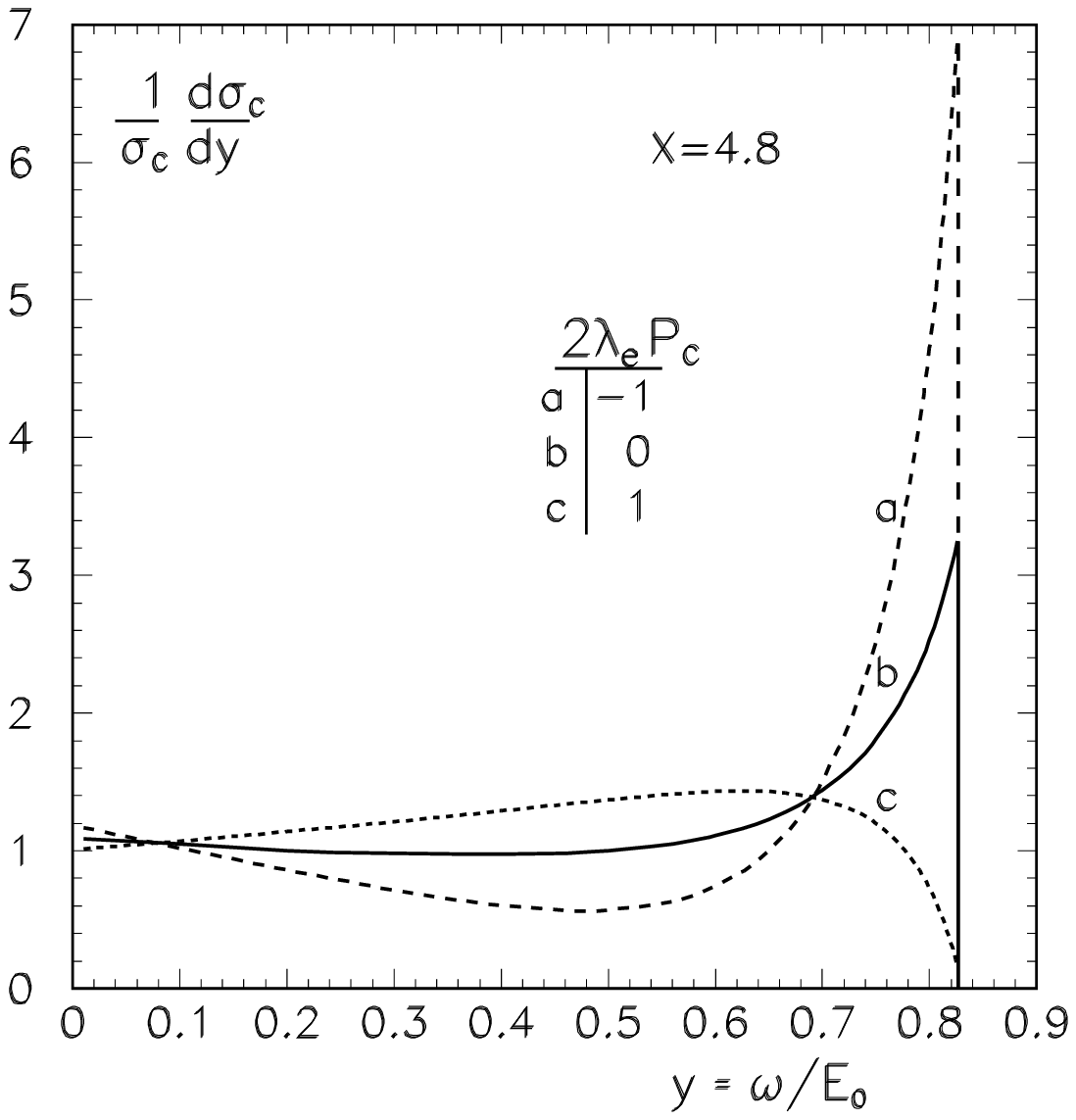}
\caption{Spectrum of the Compton-scattered photons.}
\label{f3:fig4}
\end{minipage} \hspace{0.5cm}
\begin{minipage}{0.45\linewidth}
\vspace{-0.0cm}
\hspace{-0.5cm}\includegraphics[width=7.5cm,angle=0]{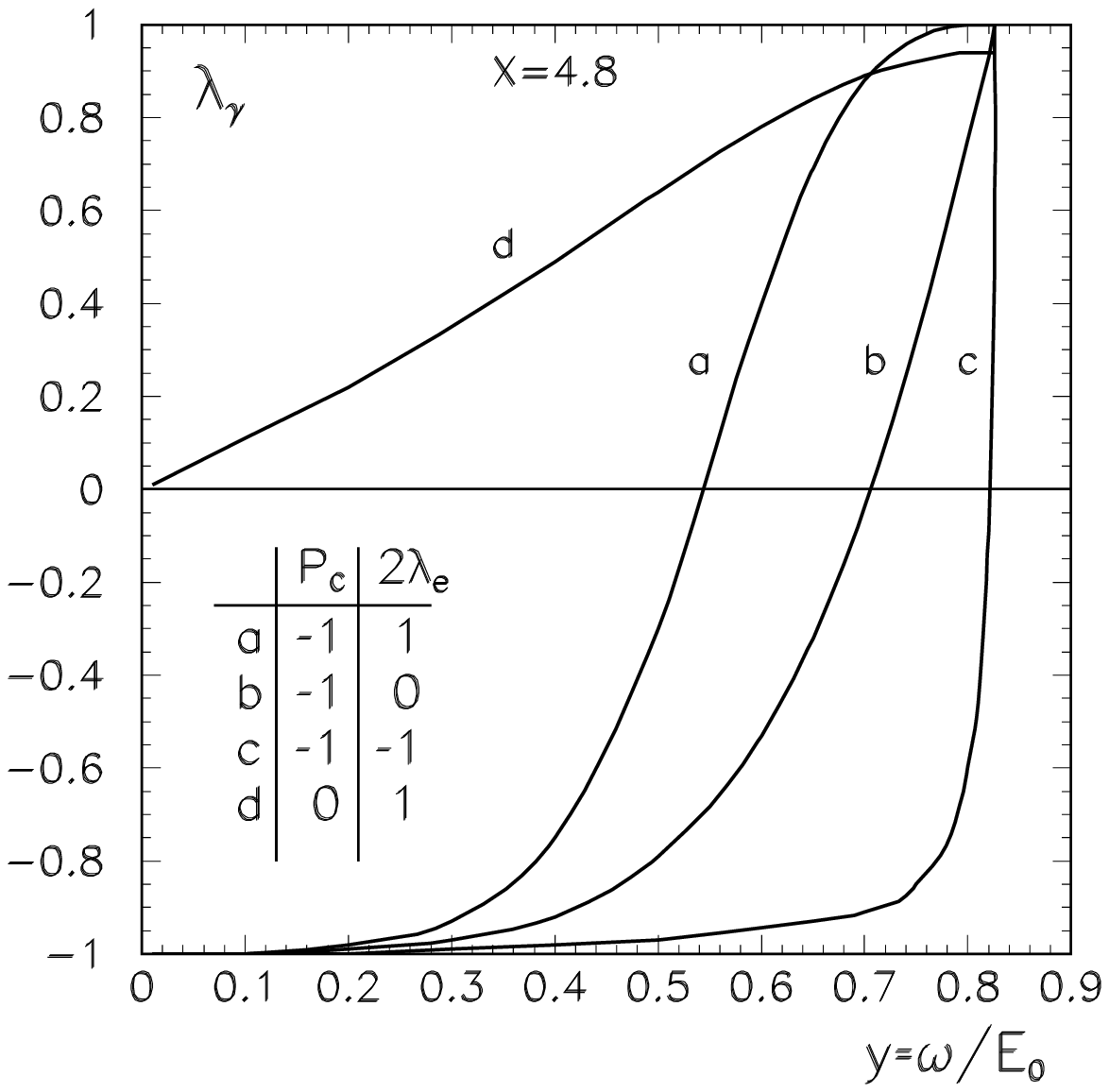}
\caption{Average helicity of the Compton-scattered photons.}
\label{f3:fig6}
\end{minipage}
\vspace{-0.cm}
\end{figure}

The energy spectrum of the scattered photons depends on the average
electron helicity $\lambda_{e}$ and that of the laser photons $P_c$.
The ``quality'' of the photon beam, i.e., the relative number of hard
photons, is improved when one uses beams with a negative value of
$\lambda_{e} P_c$.  The energy spectrum of the scattered photons for
$x=4.8$ is shown in Fig.~\ref{f3:fig4} for various helicities of the
electron and laser beams.

With increasing $x$, the energy of the backscattered photons
increases, and the energy spectrum becomes narrower.  However, at
large values of $x$, photons may be lost due to creation of \EPEM\ 
pairs in the collisions with laser photons~\cite{GKST83,TEL90,TEL95}.
The threshold of this reaction is $\omega_m \omega_0 = m^2c^4$, which
corresponds to $x=2(1+\sqrt{2})\approx 4.83$.  One can work above this
threshold, but with a reduced luminosity; the luminosity loss factor
is about 5--10 for $x=$ 10--20. Therefore, $x\approx 4.8$ is the most
preferable value. The optimum wavelength of the laser photons
corresponding to $x=4.8$ is 
\begin{equation}
\lambda= 4.2 E_0 \;[\TEV]\;\; \MKM\,. 
\label{lamb}
\end{equation}

The mean helicity of backscattered photons at $x=4.8$ is shown in
Fig.~\ref{f3:fig6} for various helicities of the electron and laser
beams.  For $2 P_c \lambda_e = -1$ (the case of the peaked energy
spectrum), all photons in the high-energy peak have a high degree of
like-sign polarization.  A high degree of circular photon polarization
is essential for the study of many physics processes.

\vspace{-0.2cm}
\subsection{Multi-photon (nonlinear) effects at the conversion region}
The electromagnetic field in the laser wave at the conversion region
is very strong, and so electrons can interact with several laser
photons simultaneously.  These nonlinear effects are characterized by
the parameter~\cite{Berestetskii,GKP,Galynskii} \vspace{-0mm}

\begin{equation}
\xi^2 = \frac{e^2\overline{F^2}\hbar^2}{m^2c^2\omega_0^2} =
\frac{2 n_{\gamma} r_e^2 \lambda}{\alpha}=0.36\left[\frac{P}{10^{18}\,\mbox{W/cm$^2$}}\right]\left[\frac{\lambda}{\MKM}\right]^2,
\label{xi2}
\end{equation} \vspace{-0mm}

\noindent where $F$ is the r.m.s.\ strength of the electric (magnetic)
field in the laser wave and $n_{\gamma}$ is the density of laser photons.
At $\xi^2 \ll 1$, the electron scatters on one laser photon, while
at $\xi^2 \gg 1$ scattering on several photons occurs.

The transverse motion of an electron through the electromagnetic wave
leads to an effective increase of the electron's mass: $m^2\rightarrow
m^2(1+\xi^2)$, and so the maximum energy of the scattered photons
decreases: $\omega_m/E_0 = x/(1+x+\xi^2)$.  At $x=4.8$, the value of
$\omega_m/E_0$ decreases by about 5\% for $\xi^2=0.3$.  For figures
demonstrating evolution of the Compton spectra as a function of
$\xi^2$ please refer to Refs.~\cite{Galynskii,TESLATDR}. With
increasing $\xi^2$, the Compton spectrum is shifted towards lower
energies, and higher harmonics appear; the part of the \GG\ luminosity
spectra that is due to nonlinear effects becomes broader. So, the
value of $\xi^2\sim 0.3$ can be taken as the limit for $x=4.8$; for
smaller values of $x$ it should be even lower. The complete set of
formulae for pair production in the laser wave for any combination of
polarizations and field strengths can be found in
Ref.\cite{Ivanov-comp}.

Nonlinear effects also exist in \EPEM\ creation at the conversion
region in collisions of laser and high-energy
photons~\cite{GKP,Galynskii1,TESLATDR,Ivanov-pair}.  There exist some
other interesting effects in the conversion region, such as the
variation of polarization of electrons~\cite{kotser} and high-energy
photons~\cite{KPerltS} in the laser wave.

\vspace{-0.2cm}
\subsection{Laser flash energy}
While calculating the required flash energy, one must take into
account the diffractive divergence of the laser beam and to keep small
the nonlinear parameter $\xi^2$. The r.m.s.  radius of the laser beam
near the conversion region depends on the distance $z$ to the focus
(along the beam) as ~\cite{GKST83}
\begin{equation} 
a_{\gamma}(z)=
a_{\gamma}(0)\sqrt{1+z^2/Z_R^2},\;\;\;\;\;a_{\gamma}(0) \equiv
\sqrt{\frac{\lambda Z_R}{2\pi}}, \vspace*{-0.mm}
\label{sLrz}
\end{equation}
where $Z_R$ is the Rayleigh length characterizing the length of
the focal region.

Neglecting multiple scattering, the dependence of the conversion
coefficient on the laser flash energy $A$ can be written as
\begin{equation} 
     k  = \frac{N_\gamma}{N_e} \sim 1-\exp \left(-\frac{A}{A_0} \right), \vspace*{-0.mm}
\label{kdef}
\end{equation}
where $A_0$ is the laser flash energy for which the thickness of the
laser target is equal to one Compton collision length. The value of
$A_0$ can be roughly estimated from the collision probability $p \sim
2 n_{\gamma}\sigma_{c}\ell = 1$, where $ n_\gamma \sim A_0/(\pi
\omega_0 a_{\gamma}^{2} \ell_\gamma )$, $\sigma_c$ is the Compton
cross section ($\sigma_c = 1.8\times10^{-25}$ cm$^2$ at $x=4.8$),
$\ell$ is the length of the region with a high photon density, which
is equal to $2Z_R=4\pi a_{\gamma}^2/\lambda$ at $Z_R \ll
\sigma_{L,z}\sim\sigma_z$ ($\sigma_z, \sigma_{L,z} $ are the r.m.s.\ 
lengths of the electron and laser bunches), and the factor 2 due to
the relative velocity of electrons and laser photons. This gives, for
$x=4.8$,
\begin{equation} 
  A_0 \sim \frac{\pi\hbar c\sigma_z}{2\sigma_c} \sim 3 \sigma_z
  [\MM],\,\mathrm{J}. 
\label{A0estimate}
\end{equation}
Note that the required flash energy decreases when the Rayleigh length
is reduced to $\sigma_z$, but it hardly changes with further
decreasing $Z_R$. This happens because the density of photons grows
but the length decreases, and as result the Compton scattering
probability remains nearly constant.  It is not helpful either to make
the radius of the laser beam at the focus smaller than $a_{\gamma}(0)
\sim \sqrt{\lambda\sigma_z/2\pi}$, which may be much larger than the
transverse electron bunch size in the conversion region. From
(\ref{A0estimate}) one can see that the flash energy $A_0$ is
proportional to the electron bunch length, and for $\sigma_z = 0.3$ mm
(ILC) it is about 1 J.  The required laser power is
\begin{equation} 
 P \sim \frac{A_0 c}{2\sigma_z} \sim \frac{\pi \hbar c^2}{4 \sigma_c} \sim 0.4
 \times 10^{12}\,W. 
\end{equation}
More precise calculations of the conversion probability in head-on
collision of an electron with a Gaussian laser beam can be found
elsewhere~\cite{GKST83,TEL90,TEL95,NLC}; they are close to the above
estimate.

However, this is not a complete picture, since one should also take
into account the following effects:

$\bullet$ {\it Nonlinear effects in Compton scattering}. The photon
density is restricted by this effect. For shorter bunches, nonlinear
effects will determine the laser flash energy.

$\bullet$ {\it Collision angle}.  If the laser and electron beams
do not collide head-on (if the laser optics is outside the electron beam), the
required laser flash energy is larger by a factor of 2--2.5.

$\bullet$ {\it Transverse size of the electron beam}.  In the
crab-crossing scheme, the electron beam is tilted, which leads to an
effective transverse beam size comparable to the optimum laser spot size.

Simulations show~\cite{TEL-Snow2005,TEL-PH05-2,Klemz2005} that if all
the above effects are taken into account, the required flash energy
for the photon collider at the ILC with $2E_0=500$ GeV and for
$\lambda=1.05$ \MKM\ is about $A \approx 9$ J, $\sigma_t \sim 1.5$ ps,
$a_{\gamma}(0) \sim 7$ \MKM. The corresponding peak power is 2.5 TW.
The optimum divergence of the laser beam is about $\pm 30$ mrad.
Lasers with $\lambda \approx 1$ \MKM\ can be used up to $2E_0 \sim
700$ GeV~\cite{TEL-Snow2005} (due to the \EPEM\ pair creation in the
conversion region).

\section{The most important advances in photon colliders}

\subsection{Early considerations, collision schemes}

In early 1980s, two linear colliders were under consideration:
VLEPP~\cite{VLEPP} and SLC~\cite{SLC}. For the photon collider, we used
the parameters presented in Table~\ref{tab1}.
\begin{table}[!htb]
\caption{Parameters of VLEPP and SLC used for \GG\ collider in 1980.}
\bc
  {\renewcommand{\arraystretch}{1.1} \setlength{\tabcolsep}{3.8mm}
\small
\begin{tabular}{l  c c} \hline
& VLEPP & SLC \\ \hline
CM energy, GeV & 200--600 & 100--140 \\
Luminosity, \CMS  & $10^{32}$ & $2\times 10^{30}$ \\
Particles in one bunch & $10^{12}$ & $2\times 10^{10}$ \\
Repetition rate & 10 & 180 \\
Trans. sizes$^*$ $a_e=\sqrt{2}\sigma_x=\sqrt{2}\sigma_y, \MKM $   & 1.25 & 1.8 \\
Bunch length, $\sigma_z$, mm  & 1.8 & 1.0 \\
Beta-function at IP, cm  & 1.0 & 0.5 \\
\end{tabular} \\
}
\ec
\label{tab1}
\vspace{-0.7cm}
\end{table}

For \EPEM\ collisions at VLEPP flat beams were considered from the
beginning. As the flatness was not necessary for \GG, for simplicity
we considered round beams with the same beam cross section. At SLC,
round beams were planned even for \EPEM. These parameters differ very
much from those in the present projects.

The second fact that influenced the initial consideration of the
photon-collider scheme was the minimum focal spot size achieved with
powerful lasers: it was about 20 \MKM, much larger than the optimum
one for a diffraction-limited laser beam.

\begin{figure}[!htb]
\centering
\vspace{-0.4cm}
\includegraphics[height=6cm, angle=0, angle=-3]{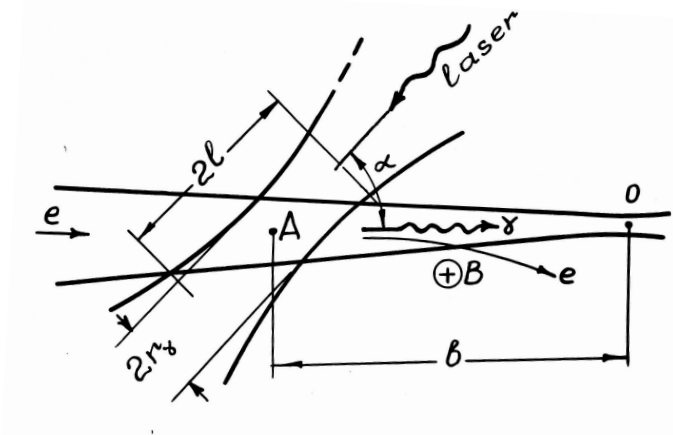}
\vspace{-0.5cm}
\caption{The scheme of the photon collider with magnetic deflection~\cite{GKST81,GKST83}.}
\label{f:scheme0}
\vspace{0mm}
\end{figure}

The originally proposed scheme of the photon collider is shown in
Fig.~\ref{f:scheme0}~\cite{GKST81,GKST83}. The laser light is focused
on the electron beam in the conversion region C, at a distance of
$b\sim 10$ cm from the interaction point O; after Compton scattering,
the high-energy photons follow along the initial electron trajectories
with a small additional angular spread $\sim 1/\gamma$, i.e., they are
focused in the interaction point O.  Electrons are swept away by a
magnetic field $B\sim 1$ T. The obtained $\gamma$ beam collides
downstream with the oppositely directed electron beam or another
$\gamma$ beam.  The required laser flash energy (for VLEPP or SLC
parameters) was about 10--20 J.

The scheme with the magnetic deflection of used beams allowed rather
clean \GG\ or \GE\ collisions to be produced. Taking $b \gg \gamma
a_e$, one can obtain a \GG\ luminosity spectrum with the width of
$\sim$10--15 \% (the ``monochromatization''
effect~\cite{GKST83,GKST84}).  The optimum distance $b$ corresponds to
the case when the size of the photon beams at the IP due to Compton
scattering is comparable to the vertical (minimum) size of the
electron beam: $b\sim \sigma_y \gamma$, that is, about $b\sim 20$ cm
for $E_0=100$ GeV and $\sigma_y=1$ \MKM.  Another factor limiting the
maximum value of $b$ is the increase of the electron beam size, which
leads to the increase of the required laser flash energy. The minimum
laser spot size attainable, 20 \MKM, allowed $b \sim 10$ cm, which was
sufficient for magnetic deflection.  Later, in 1985, the chirped pulse
amplification (CPA) laser technology emerged, which enabled production
of laser beams of ``diffraction'' quality, allowing reduction of the
spot sizes to their diffraction limits (we considered such beams as a
limiting case).

\begin{figure}[!hbt]
\centering \vspace*{-0.0cm}
\includegraphics[width=9cm,angle=0]{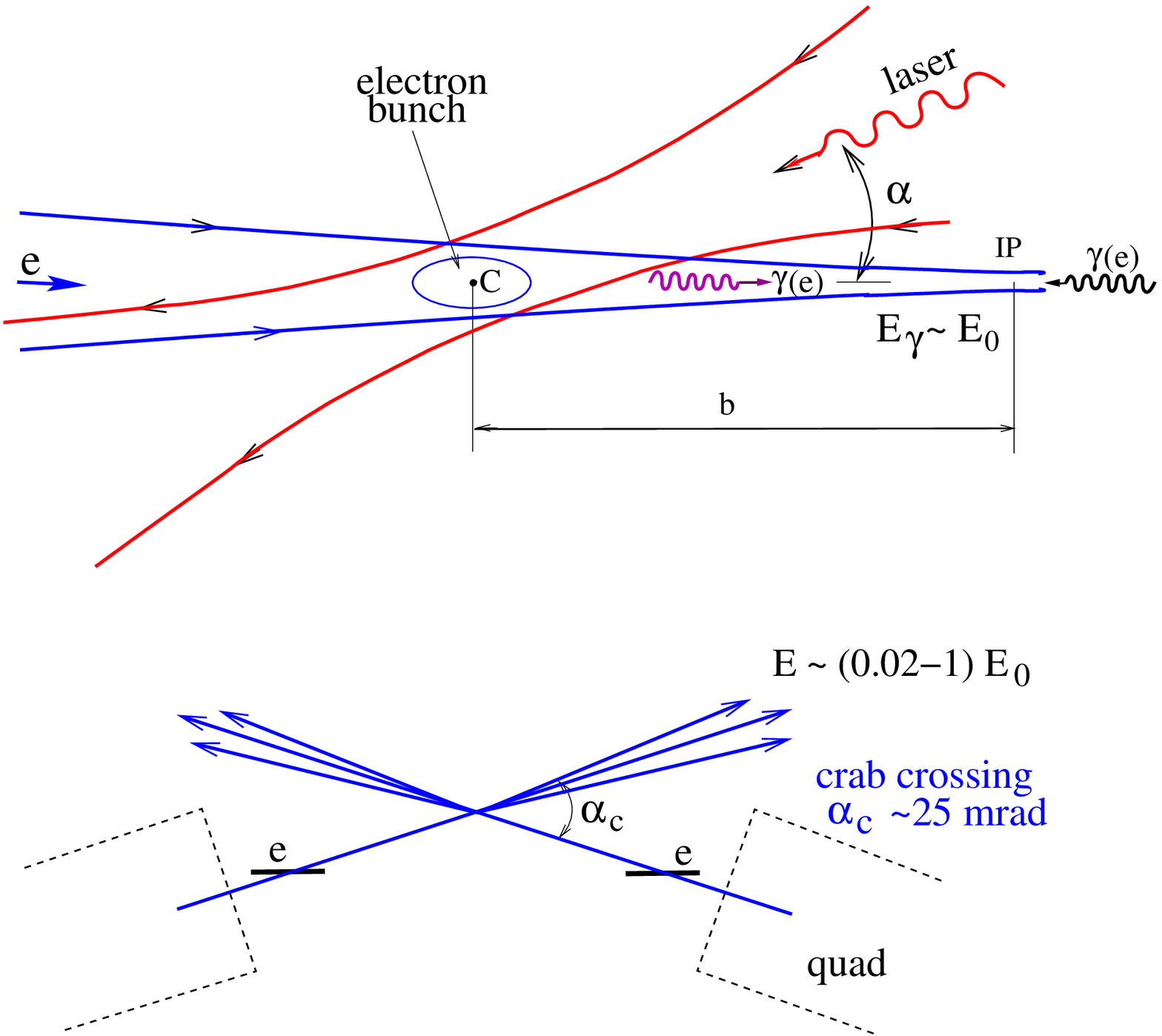}
 \vspace*{0.1cm}
\caption{Scheme of  \GG, \GE\ collider.}
\label{ggcol}
\vspace*{-0.0cm}
\end{figure}
In the following year, the vertical beam sizes in LC projects
decreased down to 3--5 nm. It became clear that on order to maximize
the \GG\ luminosity, it is necessary to focus the beam both in the
vertical and horizonal directions to the smallest possible spot cross
section $\sigma_x \sigma_y$.  Damping rings naturally produce beams
with a vertical emittance that is much smaller than the horizontal
emittance, so the resulting photon beams at the IP are flat (though
not as flat as in \EPEM\ collisons).  For $\sigma_y=3$ nm, the optimum
$b \sim \gamma \sigma_y \sim 1.5$ mm for $2E_0=500$ GeV. This space is
way too small to fit any kind of a magnet. Therefore, since
1991~\cite{TEL91}, we have been considering the scheme with no
magnetic deflection, Fig.~\ref{ggcol} (upper). In this case, there is
a mixture of \GG, \GE\ and \EMEM\ collisions, beamstrahlung photons
give a very large contribution to the \GG\ luminosity at the low and
intermediate invariant masses, the backgrounds are larger, and the
disruption angles are larger than in the scheme with magnetic
deflection. However, there are certain advantages: the scheme is
simpler, and the luminosity is larger. As for the backgrounds, they
are larger but tolerable.

Note, that even without deflecting magnets there is the beam-beam
deflection which suppress residual \EMEM\ luminosity. Also at large
CP-IP distances and a non-zero crossing angle the detector field
serves as the deflecting magnet and allows to get more or less clean
and monochromatic \GG, \GE\ collisions with reduced luminosity which
will be useful for QCD studies~\cite{TEL-mont}.

\vspace{-0.2cm}
\subsection{The removal of beams}

 After crossing the conversion region, the electrons have a very broad
energy spectrum, $E$=(0.02--1)\,$E_0$, and so the removal of such a beam
from the detector is far from obvious. In the scheme with magnetic
deflection, all charged particles travel in the horizontal plane
following the conversion.  At the IP, they get an additional kick from
the oncoming beam, also in the horizontal plane. This gave us a hope
that the beams can be removed through a horizontal slit in the final
quadrupoles; that was a feasible, but a difficult-to-implement
solution.

In 1988, R.~Palmer suggested the crab-crossing scheme for \EPEM\ 
collisions at the NLC in order to suppress the multi-bunch
instabilities~\cite{Palmer}, Fig.\ref{ggcol} (bottom).  In the
crab-crossing scheme, the beams are collided at a crossing angle,
$\alpha_c$.  In order to preserve the luminosity, the beams are tilted
by a special cavity by the angle $\alpha_c/2$.  This scheme solves the
problem of beam removal at photon colliders~\cite{TEL90}: the
disrupted beams just travel straight outside the quadrupoles.

In the scheme without magnetic deflection (which is now the primary
scheme), the disrupted beams have an angular spread of about $\pm$ 10
mrad after the IP~\cite{TESLATDR,TEL-Snow2005}. The required crossing
angle is determined by the disruption angle, the outer radius of the
final quadrupole (about 5 cm), and the distance between the first quad
and the IP (about 4 m), which gives $\alpha_c \approx 25$ mrad.

\vspace{-0.2cm}
\subsection{Luminosity}
In \EPEM\ collisions, the maximum achievable luminosity is determined
by beamstrahlung and beam instabilities. At first sight, in \GG,\GE\ 
collisions at least one of the two beams is neutral, and so the beams
do not influence each other; however, it is not so.  Beam-collision
effects at photon colliders were considered in
Refs.~\cite{TEL90,TEL95}. The only effect that restricts the \GG\ 
luminosity is the conversion of the high-energy photons into \EPEM\ 
pairs in the field of the opposing beam, that is, coherent pair
creation~\cite{ChenTel}. The threshold field for this effect $\kappa
=(E_{\gamma}/mc^2)(B/B_0) \sim 1$, where $B_0=\alpha e/r_e^2=4.4\times
10^{13}$ Gauss is the Schwinger field and $B$ is the beam field. For
\GE\ collisions, the luminosity is determined by beamstahlung,
coherent pair creation and the beam displacement during the collision.
All these processes, and a few others, were included into the software
codes for simulation of beam collisions at linear colliders by
K.~Yokoya~\cite{CAIN}, V.~Telnov~\cite{TEL95} and
D.~Schulte~\cite{Schulte}. The code \cite{TEL95} was used for
optimization of the photon colliders both at NLC~\cite{NLC} and
TESLA~\cite{TESLAgg,TESLATDR}.

It is interesting to note that at the center-of-mass energies below 
0.5--1 TeV and for electron beams that are not too short, 
the coherent pair creation is suppressed due to
the broadening and displacement of the electron beams during the
collision~\cite{TELSH,TSB2}: the beam field becomes
lower than the threshold for \EPEM\ production. So, one can
even use infinitely narrow electron beams.

\begin{figure}[!htb]
\begin{minipage}{0.5\linewidth}
  \hspace{0.5cm}\includegraphics[width=9cm,angle=0]{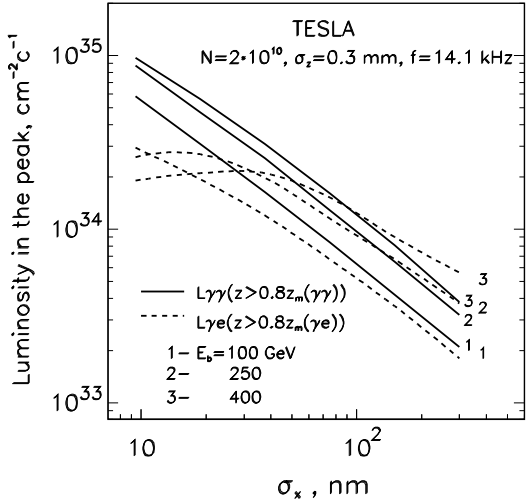}
  \vspace{-1cm}
\end{minipage} \hspace{0.5cm}
\begin{minipage}{0.4\linewidth} \hspace{3mm}
  \vspace{-4cm}\caption{Dependence of \GG\ and \GE\ luminosities in
    the high energy peak on the horizontal beam size for TESLA-ILC
    at various energies.}  \hspace{0.5cm}
\label{lumin}
\end{minipage}
\vspace{0.2cm}
\end{figure}
Simulated \GG\ and \GE\ luminosities (in the high energy peak) for
TESLA (and, similarly, for ILC) are shown in
Fig.~\ref{lumin}~\cite{TEL2001,TEL2002}. This figure shows how the
luminosity depends on the horizontal beam size. One can see that all
\GG\ luminosity curves follow their natural behavior: $L \propto
1/\sigma_x$. Note that for \EPEM, the minimum horizontal beam size
restricted by beamstrahlung is about 500 nm, while the photon collider
can work even with $\sigma_x \sim 10$ nm at $2E_0=500$ GeV, delivering
a luminosity that is several times higher than that in \EPEM\ 
collisions! In fact, the \GG\ luminosity is simply proportional to the
{\it geometric} \EMEM\ luminosity.

Unfortunately, the beam emittances in the damping-ring designs
currently under consideration do not allow beam sizes that are smaller
than $\sigma_x \sim$ 250 nm and $\sigma_y \sim 5$ nm, though a
reduction of $\sigma_x$ by a factor of two seems possible.  In
principle, one can use electron beams directly from low-emittance
photo-guns, avoiding the need for damping rings altogether, but at
present they offer a product of the transverse emittances that is
noticeably larger than can be obtained with damping rings (note: the
beams should be polarized).

To further reduce the beam emittances downstream of the damping rings
or photo-guns, one can use the method of laser cooling of the electron
beams~\cite{TSB1,Tlasv1,Tel-nano2002}. This method opens the way to
emittances that are much lower than those obtainable at damping
rings---however, this method requires a laser system that is much more
powerful than the one needed to achieve the $e \to \gamma$ conversion.
So, laser cooling of electron beams at linear colliders is a
technology for use in a \GG\ factories in the distant future.

\begin{figure}[!htb]
\vspace{-1.4cm}
\hspace{-0.0cm}\includegraphics[width=16.cm]{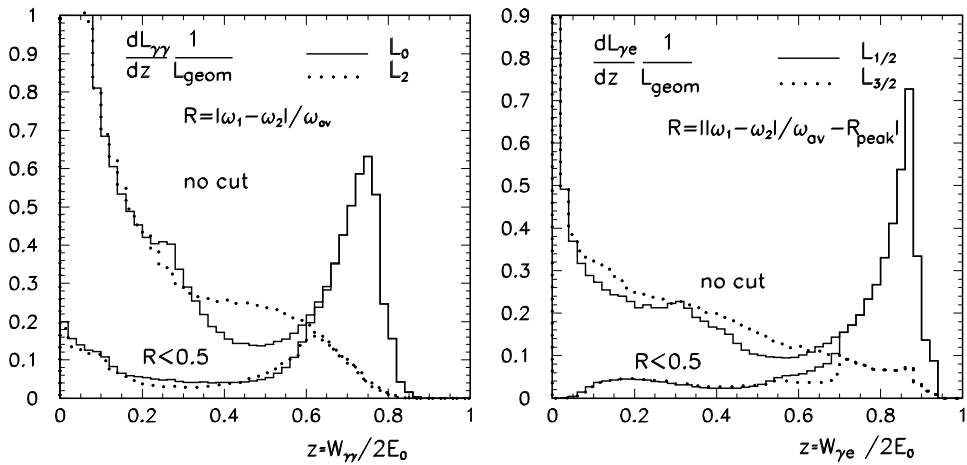}
\vspace{-1.8cm}
\caption{The \GG\ (left) and \GE\ (right) luminosity spectra for typical TESLA
  (ILC) parameters at $2E_0=500$ GeV. Solid lines for $J_z$ of two
    colliding photons equal to 0, dotted lines for $J_z=2$ (1/2 and
    3/2, respectively, in the case of \GE\ collisions). The total luminosity 
    is the sum
    of the two spectra. The residual \EMEM\ luminosity (not shown) is
    one order of magnitude smaller due to beam repulsion. }
\label{lumspectra}
\vspace{-0.2cm}
\end{figure}

The typical \GG, \GE\ luminosity spectra for the TESLA-ILC(500)
parameters are shown in Fig.~\ref{lumspectra}~\cite{TESLATDR}. They
are decomposed to states with different spins $J_z$ of the colliding
particles. The luminosity spectra and polarizations can be measured
using various QED processes~\cite{Pak,Makarenko}.
At the nominal ILC parameters (foreseen for \EPEM\ collisions), the
expected \GG\ luminosity in the high-energy peak of the luminosity
spectrum $\LGG\sim 0.15\mbox{--}0.2\, \LEPEM$~\cite{TEL-Snow2005}.  By
reducing emittances in the damping rings (which is not easy but
possible by adding wigglers), $\LGG\sim$ (0.3--0.5) \LEPEM\ can be
acieved. Note that cross sections for many interesting processes in
\GG\ collisions (e.g., charged pairs, Higgs bosons, etc) are higher
than those in \EPEM\ collisions by about one order of
magnitude~(see~\cite{TESLATDR} and references therein), so in all
cases the number of events in \GG\ collisions will be greater than in
\EPEM.

A few words about multi-TeV energies. Due to beamstrahlung, the
maximum energy of a \EPEM\ linear collider of a reasonable luminosity
is $2E_0 \sim 5$ TeV~\cite{Tel-critical}, which can be reached with
the CLIC technology. At high-energy photon colliders with short
bunches, coherent pair creation plays a role that is similar to the
role of beamstahlung in \EPEM. In the high-energy limit,
$\sigma_{\gamma}/\sigma_{e^+e^-}=3.8$~\cite{telmulti,telsnow2}, which
means that the energy reach of the photon colliders is approximately
the same as in the \EPEM\ case~\cite{TSB2,telmulti,telsnow2,telclic}.
In principle, one can imagine rather long electron bunches with a
special transverse shape, such that in the process of beam collision
the electrons are spread by the opposing beam in a more-or-less
symmetrical fashion, so that the beam field near the axis (where the
photons travel) is small, and so there is no coherent pair
creation~\cite{telmulti}. In this case, photon colliders can reach
much higher energies; alas, this is quite an unrealistic dream.

\subsection{The laser schemes and technologies}

The photon collider at the ILC(500) requires a laser system with the
following parameters (see Sect.~\ref{basics}): the flash energy $A \sim
10$ J, $\sigma_t \sim 1.5$ ps, $\lambda \sim 1$ \MKM, and the following ILC
pulse structure: 3000 bunches within a 1 ms train and 5 Hz repetition
rate for the trains, the total collision rate being 15 kHz.  
These parameters are quite similar to those discussed for VLEPP, only
the collision rate has increased by a factor of a thousand.

As has already been mentioned above, in 1981 the short-pulse Terawatt
lasers required for by a photon collider were just a dream.  A
breakthrough in laser technologies, the invention of the chirped pulse
amplification (CPA) technique~\cite{STRIC}, occurred very soon, in
1985.  In this case, ``Chirped'' means a time--frequency correlation
within the laser pulse. The main problem in obtaining short pulses was
the limitation of the peak power imposed by the nonlinear refractive
index of the medium. This limit on intensity is about 1 GW/$\CM^2$;
the CPA technique successfully overcame it.

\begin{figure}[!htb]
\centering
\includegraphics[width=12.cm,height=7.6cm,angle=0]{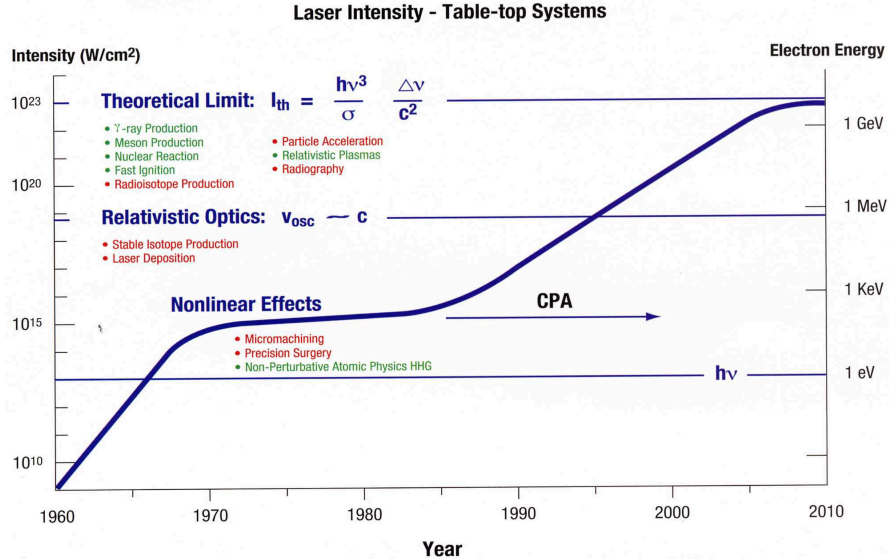}
\caption{Laser intensity vs.~year for table-top system. The progress in
  1960s and 1970s was due to $Q$-switching and mode locking;
  after 1985, owing to the chirped-pulse technique. }
\label{f-lasprog}
\end{figure}

The principle of CPA is as follows. A short, $\sim$ 100 fs low-energy
pulse is generated in an oscillator.  Then, this pulse is stretched by
a factor of $10^4$ by a pair of gratings, which introduces a delay
that is proportional to the frequency. This several-nanosecond-long
pulse is amplified, and then compressed by another pair of gratings
into a pulse of the initial (or somewhat longer) duration.  As
nonlinear effects are practically absent in the stretched pulses, the
laser pulses obtained with the CPA technique have a quality close to
the diffraction limit. This technique now allows the production of not
merely TW, but even PW laser pulses, and in several years the Exawatt
level will be reached.  Fig.~\ref{f-lasprog}~\cite{OECD} shows the
increase of the laser energy density in W/cm$^2$ vs.~year for
table-top laser systems. In 1981, the corresponded power was about
$\sim$10 GW. The minimum power required by the photon collider was
achieved roughly in 1992.

The next, very serious problem was the laser repetition rate. The
pumping efficiency of traditional flash lamps is very low; the energy
is spent mainly on heating of the laser medium. In addition, the
lifetime of flash lamps is too short, less than $10^6$ shots.
Semiconductor diode lasers solved these problems.  The efficiency of
diode laser pumping is very high, and heating of the laser medium is
low. The lifetime of the diodes is sufficient for the photon collider.

In addition to the average repetition rate, the time structure is of
great importance. The average power required of each of the two lasers
for the photon collider at the ILC is 10 J $\times$ 15000 Hz $\sim$
150 kW; however, the power within the 1 msec train is 10 J $\times
3000/0.001 \sim 30$ MW!  The cost of diodes is about ${\cal O }(1\$)
$/W, the pumping efficiency about 25\%, so the cost of just the diodes
would be at least ${\cal O }(100M\$)$, and the size of the facility
would be very large.

Fortunately, there is a solution. A 10 J laser bunch contains about
$10^{20}$ laser photons, only about $10^{11}$ of which are knocked out
in a collision with the electron bunch. So, it is natural to use the
same laser bunch multiple times. There are at least two ways to
achieve this: an optical storage ring and an external optical cavity.

In the first approach, the laser pulse is captured into a storage ring
using thin-film polarizers and Pockels
cells~\cite{NLC,TEL2001,TESLATDR}. However, due to the nonlinear
effects that exist at such powers, it is very problematic to use
Pockels cells or any other materials inside such an optical storage
ring.

\begin{figure}[!htb]
\vspace{-0.0cm}
\hspace{2cm}\includegraphics[width=13.cm]{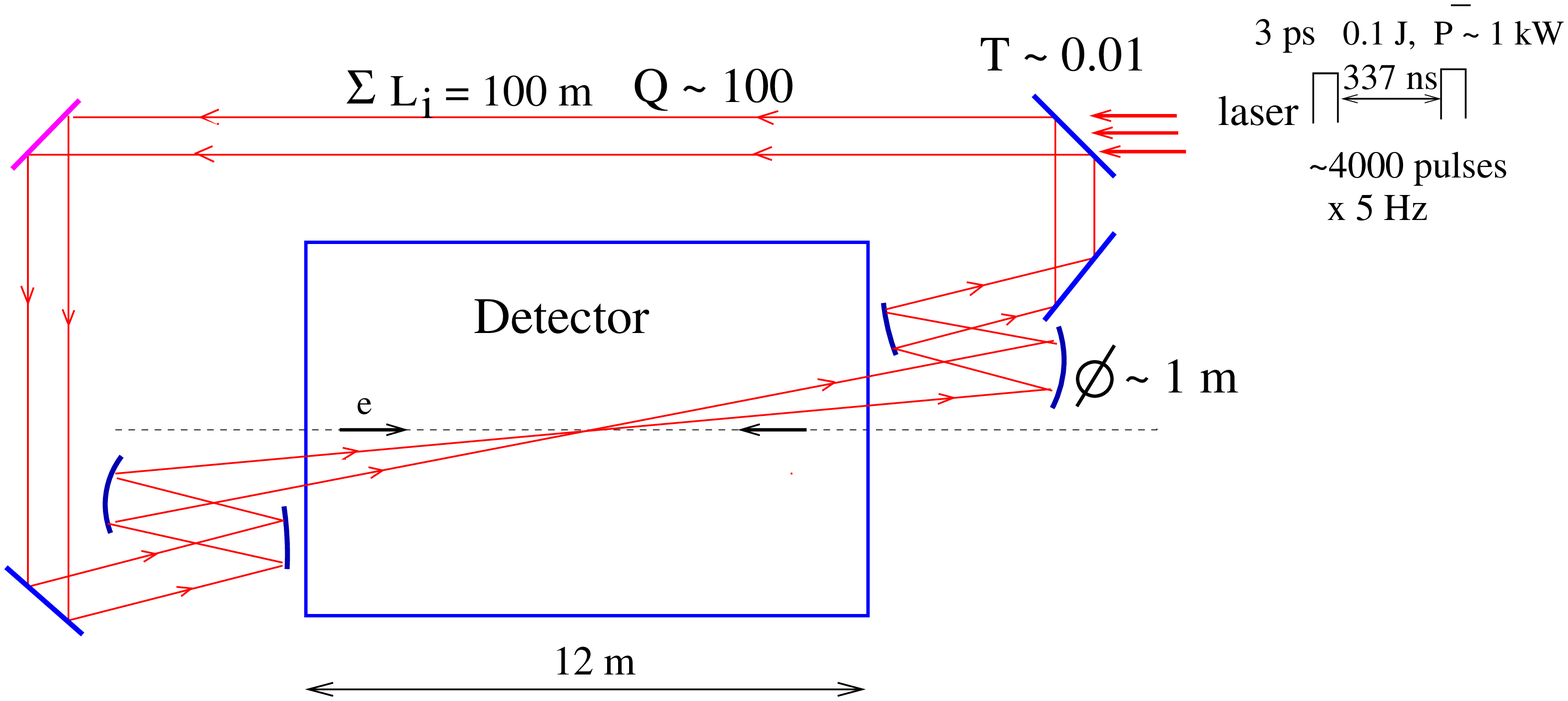}
\caption{External optical ring cavity for the photon collider}
\label{cavity}
\end{figure}

Another, more attractive approach, is an ``external'' optical cavity
that is pumped by a laser via a semi-transparent mirror.  One can
create inside such a cavity a light pulse with an intensity that is by
a factor of $Q$ (the quality factor of the cavity) greater than the
incoming laser power. The value of $Q$ achievable at such powers is
100--200. The optical-cavity principle is illustrated in
Fig.~\ref{cavity}. The cavity should also include adaptive mirrors and
other elements for diagnostic and adjustment.

While working on photon colliders, I was in contact with many laser
experts; incredibly, no one ever said a word about ``external''
optical cavities.  It was in early
1999~\cite{Hiroshima1999,Tfrei,e-e-99} that I came to the idea of such
a cavity from first principles, checked the literature, and found that
such cavities already existed, were used in a FEL experiments, in the
gravitational-wave experiment LIGO, and in the optical laboratories.
Only then did I finally come to believe in the technical feasibility
of the photon collider with TESLA-ILC pulse structure and started to
push it vigorously~\cite{TEL2001,GG2000,TESLATDR}. Working on the
TESLA TDR at DESY in 1999--2000, I got the people from the Max Born
Institute (Berlin) involved in the work on the optical cavity, and
they further advanced this scheme~\cite{Will2001,Klemz2005}; now, it
is the baseline approach for the laser system at the ILC.

Advancements in laser technologies is being driven by several large,
well-funded programs, such as inertial-confinement fusion.  This is a
very fortunate situation for photon colliders as we would benefit from
the last two decades of laser-technology developments that have cost
hundreds of millions of dollars each year. They are: the chirped-pulse
technique, diode pumping, laser materials with high
thermoconductivity, adaptive optics (deformable mirrors), disk
amplifiers with gas (helium) cooling, large Pockels cells, polarizers,
high-power and high-reflectivity multi-layer dielectric mirrors;
anti-reflection coatings, etc.  Now, practically the all laser
technologies and components required for a photon collider are in
existence; nevertheless, the construction of such a state-of-the-art
laser system would not be an easy task.

One should not forget free-electron lasers either. These might be
single-pass SASE FEL lasers or amplifiers~\cite{KONDR82,Yurkov},
though they require an excessively high electron current. More
attractive is an FEL amplifying a chirped laser pulse~\cite{TEL91}
that is then compressed by grating pairs, as in solid-state lasers. In
this case, one can use much longer electron bunches. Such FELs with
CPA were considered in Ref.~\cite{NLC,Kim} (single-pass) and in
Ref.\cite{Chan} (a multi-pass regenerative amplifier). FEL facilities
are much larger than the ``table-top'' solid-state lasers, but FELs
have certain advantages for trains with small inter-bunch spacing; in
particular, they have no problems with pumping and overheating of the
laser medium.
\section{Physics}
The \GG\ and \GE\ capabilities can be added to a high-energy \EPEM
linear collider at a small fraction of the cost of the entire project.
Although the \GG\ luminosity in the high-energy part of the spectrum
will be lower than in \EPEM\ by a factor of 3--5, the cross sections
in \GG\ collisions are typically greater by a factor of 5--10, so the
number of ``interesting'' events would surpass that in \EPEM\ 
collisions. Moreover, a further increase of the achevable \GG\ 
luminosity by up to one order of magnitude cannot be excluded.

Since the photon couples directly to all fundamental charged
particles---leptons, quarks, $W$'s, supersymmetric particles,
etc.---the photon collider provides a possibility to test every aspect
of the Standard Model, and beyond. Besides, photons can couple to
neutral particles (gluons, $Z$'s, Higgs bosons, etc.) through
charged-particle box diagrams. See S.~Brodsky's review talk at this
conference for more details~\cite{Brodsky}.

Many theorists took part in the development of the physics program for
the photon collider; the total number of publications has surpassed
the 1000 mark.

The physics program of the photon collider is very rich and
complements in an essential way the physics in \EPEM\ collisions under
any physics scenario. In \GG, \GE\ collisions, compared to \EPEM,
\vspace{-0.1cm} \bi
\item the energy is smaller only by 10--20\%; \\[-7mm]
\item the number of interesting events is similar or even greater; \\[-7mm]
\item access to higher particle masses (single resonances in $H$, $A$,
etc., in \GG, heavy charged and light neutral (SUSY, etc.) in
  \GE);  \\[-7mm]
\item at some SUSY parameters, heavy $H/A$-bosons will be seen only in
\GG;\\[-7mm]
\item higher precisions for some phenomena; \\[-7mm]
\item different types of reactions;  \\[-7mm]
\item highly polarized photons.  \ei

  So, the physics reach of a \GG, \GE\ and \EPEM\ colliders is
  comparable. The only advantage of \EPEM\ collisions is the narrower
  luminosity spectrum, the feature that is of rather limited use.
  
  The photon collider can be added to the linear \EPEM\ collider at a
  very small incremental cost. The laser system and modification of
  the IP and one of the detectors would add about 3--4\% to the total
  ILC cost.  Some decrease of the \EPEM\ running time is a negligible
  price to pay for the opportunity to look for new phenomena in other
  types of interactions.
  
  More about physics at \GG colliders can be found in reviews
  \cite{Brodsky,TESLATDR,Boos,Velasco,DeRoeck,Krawczyk,Brodsky1,Zerwas},
  references therein, and many other papers.
\section{Studies, projects, politics}
Photon colliders were discussed at the series of LC, LCWS and PHOTON
workshops/conferences, and at many others. In the beginning, these
were single talks, then working groups formed, and then International
Workshops on Photon Colliders took place in Berkeley in 1994
(A.~Sessler)~\cite{BERK}; at DESY in 2000 (R.~Heuer,
V.~Telnov)~\cite{GG2000}; at FNAL in 2001 (M.~Velasco)~\cite{GG2001};
in Warsaw, 2005 (M.~Krawczyk)~\cite{GG2005}, as well many smaller
meetings.

Several LC projects have been in existence: VLEPP, NLC, JLC, TESLA,
SBLC, CLIC---and each one of them foresaw the \GG, \GE\ option. In
1996--1997, three LC projects published their Conceptual Designs with
chapters or appendices describing a second IP, dedicated for a photon
collider: NLC~\cite{NLC} (ed. K.-J. Kim), TESLA/SBLC~\cite{TESLAgg}
(ed. V.~Telnov), JLC~\cite{JLC,JLCgg} (ed. T.~Takahashi, I.~Watanabe).
In February 1999, at the \GG\ mini-workshop on photon colliders in
Hiroshima, it was decided to organize an International Collaboration
on Photon Colliders. This was announced at LCWS1999, and approximately
150 physicists signed up. The work was done, presumably, within the
framework of regional studies.

All that time, photon colliders were considered first and foremost as
a natural additions to the \EPEM\ collider projects. However, there
were several short-lived suggestions to build dedicated photon
colliders with no \EPEM\ collisions at all: a 10 GeV \GG\ collider for
study of $b$-quark resonances~\cite{Bauer}, a 100--200 GeV \GG\ 
collider for ``Higgs hunting''~\cite{Bal-Gin}, a
``proof-of-principle'' photon collider at the SLC~\cite{Gron}, a
photon collider on the basis of the CLIC test facility~\cite{Asner}.
In my mind, suggesting a linear collider with no \EPEM\ collisions
when most people dream about \EPEM\ is just not serious. ``Test''
colliders with low energy or luminosity would be a waste of resources.

A few words about dedicated \GG\ workshops. At LC92, I spoke to Andy
Sessler about photon colliders and asked him to give a talk on the
possible application of FELs for photon colliders. He did so, and ``in
addition'' organized the first workshop on \GG\ colliders (Berkeley,
1994 ~\cite{BERK}), gave a talk on photon colliders at PAC95, and wrote
a paper on photon colliders for Physics Today~\cite{Sessler}.

The second International Workshop on the High Energy Photon
Colliders~\cite{GG2000} (GG2000) was organized at DESY as a part of
work on the Photon Collider for the TESLA TDR~\cite{TESLATDR}.
Together with accelerator physicists, we found that after some
optimization of the damping rings and the final focus system,
$\LGG(z>0.8z_m) \sim 0.3 \LEPEM$ can be achieved. Now, even some of
the past opponents of photon colliders agreed that \GG, \GE\ should be
built. As for the technical feasibility, the very attractive idea of
an external optical cavity was already in existence in 2000.

The primary motivation behind GG2001~\cite{GG2001} at FNAL was the
idea of $e\to \gamma$ conversion using crystals instead of lasers.  It
was rejected, completely and outright, due to the destruction of
crystals by the very dense electron beams, large photo-nuclear
backgrounds and defocusing by the beam produced plasma; this was quite
obvious from the very beginning~\cite{VLEPP1980,TEL90}.

Now, let me discuss the present situation.  Due to the high costs of
building a high-energy linear collider, the international HEP
community agreed to build one collider for the energy $2E_0=0.5$--1
TeV instead of three (TESLA, NLC and JLC). In 2004, the ILC project,
based on the superconducting TESLA-like technology, was inaugurated.
According to the consensus document titled ``Understanding matter:
...the case for the Linear Collider''~\cite{docum}, which was signed
by three thousands supporters, the ILC should have an interaction
region compatible with the photon collider. So, the next steps are the
ILC design, cite selection, obtaing government approval and funding,
and the construction.  Under the best-case scenario, ILC operation may
start in 2015.  ``To be or not to be'' for the sub-TeV linear collider
depends both on the energy scale of new physics, which should become
known soon after the start of experiments at the LHC, and on multiple
other scientific and political factors.

If the ILC is built, in the first few years it will operate in the
\EPEM\ mode in all (1 or 2) of its detectors.  Then, one of the IPs
would be modified for the \GG, \GE\ mode of operation.

Unfortunately, life is not easy for the advocates and supporters of
the photon collider at the ILC. For many years, the photon collider
has been considered an ``option'' in the``baseline'' \EPEM\ collider
design. In real life, ``option'' meant no support, and no money.
Nevertheless, the interest of the physics community in having the
photon collider built is tremendous.  For example, the number of
articles in the SPIRES database (publications only) that mention
linear colliders or photon (gamma) colliders in their titles are,
respectively, approximately 2950 and 600. These numbers speak for
themselves.

In the conclusion of my sermon, let me share with you an instructive
story from the very early days of collider physics.  When G.~Budker
proposed to build the first \EPEM\ storage ring in Novosibirsk,
responses of all three referees were negative. However, I.~Kurchatov,
head of the USSR nuclear program, overruled the skeptics: "If the
referees are so unanimously against it, it means that the project is
really interesting"---and gave the green light to \EPEM\ colliding
beams. So, nothing new is ever easily done.

  In summary: the physics expected in the 0.1--1 TeV region is very exciting, 
there is a big chance that a linear collider will be built somewhere in the world, 
and then the photon collider will inevitably happen.
  \section*{Acknowledgements}
  I am very grateful to D.~Asner, V.~Balakin, T.~Barklow, K.~van
  Bibber, R.~Brinkmann, S.~Brodsky, D.~Burke, H.~Burkhardt,
  S.~Chattopadhyay, P.~Chen, J.P.~Delahaye, A.~De~Roeck, K.~Geissler,
  I.~Ginzburg, J.~Gronberg, A.~Finch, R.~Heuer, C.~Heusch, G.~Jikia,
  K-J.Kim, G.~Kotkin, M.~Krawczyk, D.~Meyerhofer, D.~Miller,
  G.~Mourou, V.~Serbo, K.~M\"onig, O.~Napoli, R.~Palmer, M.~Peskin,
  F.~Richard, B.~Richter, A.~Seryi, A.~Sessler, S.~S\"oldner-Rembold,
  J.~Spenser, A.~Skrinsky, T.~Takahashi, T.~Tauchi, D.~Trines,
  A.~Undrus, N.~Walker, I.~Watanabe, K.~Yokoya, P.~Zerwas for useful
  discussions and joint efforts in the study of photon colliders, and
  to all the people who have contributed to the development of the
   photon collider and its physics program.

\end{document}